# Strategies to Avoid Illegal Data Access

Muhammad Mubeen[1], Muhammad Arslan[1], Giri Anandhi[2,]*

**Abstract**

*For companies of all sizes, data security is a top priority. The chance of unauthorized data access increases as technology develops. To prevent unwanted access to their data, businesses must be proactive. This study examines technology solutions, personnel training, and policy enforcement as methods to prevent unauthorized data access. Data may be protected from illegal access using technological solutions like firewalls, intrusion detection systems, and encryption. Intrusion detection systems notify the administrator when suspicious behavior is found, while firewalls serve as a protective border between the internal network and the internet. Even if data is intercepted, encryption makes sure it is safe. Another effective method of avoiding unauthorized data access is employee education. Employees must be taught how to spot hazards like phishing emails and shady websites and react to them. Additionally, they should be taught the right way to utilize passwords and other security precautions. To secure data, organizations should create and implement policies. Policies should set out appropriate data and system use guidelines and provide repercussions for noncompliance. Policies should be evaluated regularly to ensure that they are current and useful. Businesses may prevent unwanted access to their data by installing technology solutions, training staff, and enforcing regulations. Organizations may reduce data breach risk and maintain regulatory compliance by taking these precautions. However, several cyber threats that corporations currently face are examined in this study. It outlines several threat categories, including ransomware, phishing, malware, and data breaches, and how they might affect enterprises. It also looks at methods businesses can employ to reduce the danger posed by these threats, including setting up firewalls and antivirus software, attending frequent security awareness training sessions, and implementing robust authentication systems.*

**Keywords:** Software, data, encrypt data, malware threat, SIEM, network traffic, multifactor authentication

## INTRODUCTION

Today, enterprises all around the globe have data protection as one of their top priorities. Only InfoSec, and information security, can provide data security since it mainly focuses on preventing unwanted access to information. Soon, it is anticipated that most cyber security investments will move toward managed security services like security information and event management (SIEM). Companies are prioritizing cyber security and establishing stronger and more reliable security procedures to thwart intrusions from outsiders or nefarious insiders. Is the data protected against illegal access? Let us look at some security-related actions and strategies that help avoid the leakage of an organization's data from the hands of unauthorized personnel. Organizations should implement several high-level security methodologies to safeguard their data from unwanted access. Here are some suggestions to assist in guarding against unwanted data access.

***Author for Correspondence**
Giri Anandhi
E-mail: giri_anandhi@hotmail.com

[1]Scholar, Department of Computer Science, University of the People, Pasadena, California, USA
[2]Associate Professor, Department of Computer Science, University of the People, Pasadena, California, USA

## METHODOLOGIES TO SAFEGUARD DATA FROM UNWANTED ACCESS

### Keep Current on all Security Patches

Keeping up with all security patch updates is one of the tactics. Every company should keep its security updates current to avoid unauthorized data access. Why are that drivers, Software, software platforms, and other aspects that hackers may exploit to access the device and the user's data fixed through security upgrades? Security updates for Android, Linux, Windows, and iOS are important because operating system flaws might have detrimental effects. When recapitalization becomes possible, update Software and drivers often. The WannaCry virus was among the deadliest attacks in recent memory [1]. It was launched using the Eternal Blue exploit and exploited the Windows SMB V1 (Server Message Block) protocol vulnerability.

It is noteworthy to note that security updates for these flaws were accessible far in advance of the assault. However, thousands of people were attacked because they had not upgraded their security updates. Users may have avoided allowing illegal entry for both system assaults by employing the most current security upgrades. In order to defend the operating system and other Software from attacks, be sure to get the most recent security patches and updates. For the machine to download and apply any security fixes or improvements made available, one may also enable automatic updates. By being knowledgeable and ready, the user can protect the data from anyone seeking to access it without authorization.

### Detect and Respond to Intrusions Quickly

The other tactic is to identify and respond to intruders swiftly; obviously, one would need to be prepared and vigilant to protect hackers from obtaining unauthorized access to data. Nevertheless, what if the individual failed to detect an intrusion? Next, what should we do? It could be dealt with more quickly if one notices an intrusion early enough. Although prevention is essential, monitoring user behavior, connection attempts, logs, and other operations may also provide important details about the system's security. There are numerous techniques to identify and react to intrusions immediately: IDS/IPS stands for intrusion detection and prevention system. Using behavioral heuristics or known intrusion signals, an IDS examines network traffic for possible activity [2]. Intrusion detection is watching and analyzing network or system behavior for possible indicators of intrusion occurrences, such as looming policy violations, threats. However, proactive IPS surveillance of incoming traffic aids an IDS in identifying fraudulent requests.

By banning unsuitable or offensive websites, forbidding dangerous information, and warning security staff of possible security threats, an IPS stops intrusion assaults. Using a Security Issue Event Manager, security administrators may learn more about what is happening in an IT environment. The IT infrastructure of the company's operation, including its networks, host systems, programs, and security devices, produces log data, which is gathered and examined by SIEM software.

The program then finds occurrences and incidents, classifies them, and analyzes them. Tracking records of incidents and security-related events, such as attempted and successful malware activity, login attempts, or any other suspicious behavior, is one of SIEM's two main objectives to produce reports on these events and incidents. Any unusual action that suggests a security danger should be reported to security officers. Implement behavioral analytics for users and events (UEBA). The user must master analytics if the user wants to avoid unwanted data access. Analyzing client and event behavior enables the identification of any instances of aberrant behavior or departures from users' “usual” behavioral tendencies. For instance, the system would quickly inform the administrator if a user suddenly began downloading terabytes of data while typically downloading files day after day. By identifying deviations from established patterns using algorithms, statistical analysis, and computer vision, user and event behavioral analytics may identify which anomalies are occurring and how they can pose a hazard [3]. An individual will be informed of any unlawful data access in this manner. Such analytics primarily concentrate on the individuals and entities that make up a system, notably internal threats like workers who could abuse their authority to launch specialized attacks or fraudulent endeavors.

**Implement the Principle of Least Privilege (Minimize Data Access)**

One can also apply the security principle called the Principle of Least Privilege (PoLP) states that users and processes should only be given the minimal amount of access required to carry out their functions. Any security plan should include PoLP since it prevents unwanted access to sensitive information and resources. Giving a user or process just the rights they need to carry out their work is the aim of PoLP implementation. Users and processes are given roles, and privileges are given depending on those roles to accomplish this. In contrast to a user with a primary user role, one with an administrative position would have additional rights. A smaller attack surface and unauthorized information access are avoided by restricting the rights allowed. Users must only utilize the rights they have been given, according to PoLP. Users should only attempt to access resources or content within the bounds of their responsibilities. Organizations can ensure that only authorized users get access to sensitive information and resources by imposing PoLP. Policies may be used to impose PoLP, such as the requirement that users log in before accessing sensitive information [4]. Access control lists may also guarantee that only specific users can access particular resources. In order to hinder unwanted access, data encryption may also be utilized. Lastly, companies should frequently check user access and verify logins to make sure PoLP is just being followed. In order to make sure that access to outdated and superfluous data and resources is not permitted, organizations should assess the rights given to users and processes. Employing PoLP will help companies lower the risk of illegal access to information and resources [1]. Organizations may ensure that only approved users have access to critical data and resources by restricting the rights issued and keeping track of user access.

**The Use Multi-Factor Authentication**

Authenticate using several factors. In contrast to multi-factor authentication, businesses must use robust authentication by enforcing stringent password guidelines. That may significantly reduce the likelihood of illegal data access. Multifactor, as the term implies, many data must be submitted by the user in order for the system to verify it before granting access. Consequently, it is more difficult for attackers to access user accounts than for them to guess the password [3]. Examples of multifactor authentication include setting a security question, using a one-time password sent over that out channel of communication, such as an automatic phone call or text message to the user's authorized device, or biometric identification. The verification process becomes more challenging, but the security is greater since the attacker must compromise both the password and the second factor in conjunction with the first. This makes it far more challenging for an intruder to compromise authentication. They were interested in learning how to prevent unauthorized access to the data.

**Employ passphrases**

Users may choose to employ passphrases rather than passwords, even if multifactor authentication is a good idea. An example of a catchphrase, a string of unrelated words, or a sentence with spaces between the syllables is “the bundle password”. Any combination of unrelated words and symbols may be used as a passphrase; it does not need to be perfectly correct. Complicated passphrases might make complex passwords easier to remember. The generation of strong passphrases still requires caution. Even simple passphrases made up entirely of words from a common language may be broken [4].

**Implement IP Whitelisting**

IP whitelisting is a technique and a strategy for limiting access to a network or system to just those IP addresses that have been permitted. Resources are shielded from illegal access while being accessible securely. A list of IP addresses authorized to access the contents is created to do this. Requests from IP addresses not included in the list will be banned or rejected. Only authorized users should have access to the data, and IP whitelisting is a reliable approach to guarantee this [5]. Many businesses, including banks, hospitals, and governmental entities, utilize it to prevent unauthorized individuals from accessing sensitive data. Networks, web servers, and databases are just a few examples of the resources that may be controlled by businesses utilizing IP whitelisting. This stops individuals who are not authorized from accessing data.

IP whitelisting may be used to stop malicious activity on a system or network. Organizations may make sure that corrupt individuals cannot access critical data by restricting access to the system to certain IP addresses. This will make the system more resistant to harmful assaults such as ransomware and viruses. Access to business networks is also managed through IP whitelisting. Organizations may guarantee that only approved users obtain access to the network by restricting access to specified IP addresses. This aids in stopping unauthorized users from using the network and obtaining private information. As a result, IP whitelisting is a successful method for restricting the use of networks and resources and preventing illegal access to critical data. It may aid in safeguarding business networks from harmful assaults and defending enterprises against them. Companies may guarantee that company data is safe and shielded from illegal access by employing IP whitelisting.

**Encrypt Network Traffic Inside the System**

Another strategy for preventing unwanted access to data within the system is encrypting network traffic. SSH, SSL/TLS, and IPsec are a few techniques that may be used to do this. These enable data encryption so that only the intended receiver may decrypt it [6]. Using the network protocol IPsec, two endpoints may communicate securely. It identifies the sender and recipient and encrypts the data packets delivered across the network. In addition, IPsec permits the usage of encryption keys that may be used for data encryption and authentication. By doing this, it is made sure that the information is kept safe and only the designated user may access it.

The SSL/TLS protocol protects data while it is being sent. In order to keep the data confidential, it employs encryption and offers a secure link between the two endpoints. The encryption encrypts the data while it is in transit and authenticates the sender and recipient using a mix of private and public keys [7]. Data may be securely sent between two destinations thanks to the secure communication protocol SSH. It establishes an encrypted connection between both endpoints, enabling safe data transfer. Encryption keys can be used to verify and encrypt data, and SSH allows their usage. Data may be easily protected from unwanted access by encrypting network traffic within the system. Data may be kept safe and only available to the intended receiver by employing IPsec, SSL/TLS, and SSH. This enables private data transmission while enabling a secure connection between two endpoints.

**Encrypt Data-at-Rest**

Sensitive information on a computer or server is safeguarded against unwanted access using data-at-rest encryption. Data-at-rest encryption guarantees that information is encrypted, rendering it inaccessible to anybody without the necessary encryption key [8]. Businesses can stop unwanted access to sensitive information and safeguard it from harmful actors by encrypting data at rest. For data to be decoded by someone with the right key, it must first be encoded into an unintelligible form. Typically, this key consists of a random string of numbers and characters that only the data's sender and recipient know [5]. Data-at-rest that has been encrypted can only be read once the key has been used to decode it. The data remains illegible even if a third-party gains access to the system without the proper key.

Any company's security architecture must include data-at-rest encryption as a key element. Limiting access and guarding against data breaches aids in maintaining the security of sensitive data. It also conforms with several laws and rules, like GDPR and HIPAA, which demand that enterprises preserve sensitive data. Encrypting data while it is at rest may also assist enterprises in preventing theft or exploitation of their intellectual property.

Implementing software or hardware encryption techniques is the norm for data-at-rest encryption. Software encryption options are often employed when it comes to encrypting specific files or folders, while robust encryption solutions are used to secure whole hard drives or storage devices. As an additional measure of protection, several businesses combine hardware and software encryption solutions [9]. Any firm should consider encryption as a crucial security tool. Preventing unwanted access to sensitive information helps keep that data safe and in compliance with numerous laws and regulations. It also helps businesses in defending their intellectual property and avoids data breaches.

**Ensure Anti-Malware Protection/Application Whitelisting**

The other strategy is to make sure your computer has anti-malware Software. Allow listing is a tactic that aids businesses in defending their systems and data against unwanted activity. It serves as a security mechanism to stop illegal programs from operating on a system. Since this tactic is based on the whitelisting concept, only programs that the system administrators have authorized may be run. The tactic entails compiling a list of official apps that may run on the system; any other software that does not appear on the list is not permitted to be executed [10]. This aids in preventing harmful actions that an unapproved program could start. Malicious Software, for instance, may be used to send spam emails, transmit viruses, and obtain unauthorized access to sensitive information. Organizations may reduce the potential hazards brought on by harmful actions by limiting the programs permitted to execute.

Organizations may also ensure the accuracy of the data saved on the system by implementing an application whitelisting policy. The ability to only run apps that have been allowed by the system administrators helps to guard against unauthorized access to the information. By doing this, the data is shielded from harmful Software, and only permitted users can access it. Additionally, the system's speed is enhanced by the application whitelisting approach. The system saves resources by only processing and executing authorized applications since only those approved may be performed. This enhances the system's overall functionality [5]. In general, the Ensure Anti-Malware Protection/Application Safelisting approach assists businesses in defending their data and systems against harmful attacks and illegal access. Restricting the number of apps that are permitted to execute also aids in enhancing system performance.

**Track and Manage Your Risks**

Using the “Track and Manage Your Risks technique”, businesses may prevent unauthorized data access. It entails comprehending the dangers caused by unauthorized access to information and creating plans to lessen or eliminate those dangers. Finding risk sources is the initial stage of this method. This entails determining the kind of data being held, the organization's existing security posture, and any possible hazards related to the data. For instance, a company that stores personal data must consider the likelihood of a security breach or other unlawful access. The company may develop a risk management strategy after recognizing the hazards.

The risk management strategy should include safeguards against unauthorized access to the data. Strong access restrictions, including passwords, encryption, and two-factor authentication, might be used as one of these precautions. A response strategy should be in place, and the business should also keep an eye out for any possible security problems [7]. The company should establish guidelines and practices for managing data. It also covers the rules for who may access and use the data and how it is kept. To guarantee data security, these principles should be updated and reviewed often. Finally, organizations need to educate their staff members on data security and the significance of preventing illegal access to data. This entails educating staff members on the company's security policies and practices and regularly reminding them of the significance of data security. In general, the method of “Track and Manage Your Risks” may assist enterprises in lowering the risk of unauthorized data access. Organizations may ensure that data is safe and shielded from unauthorized access by comprehending the dangers of unauthorized access, creating a risk management strategy, and training personnel.

**The use of the Firewalls**

The implementation of firewalls is a different technique. A firewall is a reliable gatekeeper. It records attempts to enter the operating system and halts prohibited or undesired activities. How is this done, exactly? A firewall, which functions as a screen or barrier, keeps the computer and also another network, such as the internet, apart. A traffic controller and a firewall could be compared. The user network and the client's data may be secured by managing network traffic. Incoming network traffic not requested has to be filtered away, and access is confirmed by checking it for spyware and other hazardous network traffic. A firewall is often pre-installed on the operating system and the security application. Check to see whether such functions are turned on.

Additionally, ensure the security settings are set up to execute updates automatically by checking their settings. A firewalled system initially analyzes network traffic using rules. Only inbound connections configured to be allowed by a firewall are accepted. It does this by determining, following previously established security standards, whether to accept or reject a series of data packets—the communication units that transfer over digital networks. A firewall serves as a traffic guard at the computer's port or entry point. Only reliable sources or IP addresses are allowed. Due to their ability to identify a machine or source, IP addresses are important.

The user's residence is revealed via their postal address. Another tactic that, when used correctly, has a good probability of preventing unwanted employees from accessing data is biological surveillance. Monitoring user behavior helps to protect data while ensuring its accessibility and adherence to data privacy and security requirements. UAM performs more tasks than only monitoring network activity. Instead, it can monitor every user's action, such as how they browse the internet and if they access private or illegal data. All user-performed systems, data, applications, and network actions fall under this category.

All information obtained must be examined within the constraints of corporate policy and the user role to determine if improper conduct occurs. The organization using the UAM system determines what “inappropriate user behavior” is. It might encompass everything from browsing personal websites or shopping during business hours to stealing confidential corporate information like intellectual property. Large volumes of data may be accumulated at every monitoring level [10]. Any user behavior monitoring tool should seek out and eliminate information that might be used to further data protection efforts. Organizations can quickly identify and look into questionable user behavior with efficient procedures in place. Additionally, the user may discover if individuals are using unapproved services and apps, transferring private data to public clouds, or participating in other harmful activities when accessing the network or other resources of the business. Tools for tracking user behavior are also useful for preventing workers from sharing private information with them when they leave the organization.

**The Physical Monitoring**

Large volumes of data may be accumulated at every monitoring level. Whatever user activity monitoring tool should seek out and eliminate information that might be used to further data protection efforts. Organizations can quickly identify and look into questionable user behavior with efficient procedures in place. Additionally, management may discover if individuals are using unapproved services and apps, transferring private data to public clouds, or participating in other harmful activities when accessing the network or other resources of the business [8]. Tools for tracking user behavior are also useful for preventing workers from taking any private information with them when they leave the organization, either financial or real estate data.

**Summary**

As more data is kept on computers and in the cloud, data security is an increasing problem for companies and organizations. Organizations must ensure that their data is protected from unwanted access, given the growing volume of data being kept online. Using tactics to prevent unauthorized data access is one of the best methods to secure data. Ensuring that almost all data is kept securely is among the most crucial measures to prevent unauthorized data access. All data should be encrypted, and organizations should take precautions to guarantee that only vetted employees may access it. In order to promptly recover any lost or stolen data, enterprises need a solid backup system in place. Organizations should also put rules in place that prohibit sharing sensitive information and mandate that users only be granted access to the data they need to do their duties. In order to spot any unwanted access, organizations also should take precautions to guarantee that their systems are routinely monitored. Implementing technologies like intrusion detection or remote monitoring software may help with this. Organizations should also conduct regular audits to identify security system weaknesses or vulnerabilities. Organizations should also use the most recent security technology to safeguard their data from cyberattacks, such as firewalls and antivirus software. Organizations should also educate their workers on proper security procedures, such as using strong passwords and the confidentiality of information.

Lastly, organizations must ensure that sensitive data is frequently backed up and kept in safe offsite places. This will make it possible to retrieve any lost or stolen data rapidly. By following these actions, organizations may guarantee that sensitive data is safe and shielded from unwanted access. By implementing these measures, organizations better safeguard their data and prevent unauthorized access.

## CYBER THREATS

In the digital era, cyber dangers are of increasing concern to enterprises. Cyber threats are vicious assaults that employ networks or computer systems to obtain sensitive data, impede business, or inflict other damage. Hackers, criminals, nation-states, and even employees may threaten the internet. Organizations may encounter cyber dangers, including malware, ransomware, phishing, and social engineering, among other threats.

Sony is among the most well-known companies to experience cyber-attacks. A gang of hackers known as the “LulzSec” organization breached Sony's networks. Millions of Sony consumers had their personal information exposed due to this assault. The corporation suffered a severe financial setback due to the assault, and it took Sony many months to bounce back. Target is another well-known company. Hackers broke into Target's computer systems and stole millions of customers’ personal information [11]. Target suffered significant financial losses as a result of the hack, and they were forced to compensate millions of consumers. Yahoo also had a large data leak. This hack was the biggest data breach ever, compromising the personal information of individuals. To compensate its consumers whom the leak had impacted, Yahoo made a sizable payment. This assault significantly impacted Yahoo, and it took the business many months to recover. The above are some of the organizations that have faced cyber threats among many organizations as threats happen to date. Organizations may help safeguard themselves against cyber threats and the potential harm they inflict by adopting some proactive measures. Here are examples of certain cyber threats that affect organizations.

### Malware Threat

A kind of destructive Software called malware is created to infiltrate and harm computer systems. It may spread through websites, email attachments, and security flaws, among other methods. Malware can steal private data, interfere with regular system functions, and even cause hardware damage. Cyberattacks like distributed denial-of-service (DDoS) and denial-of-service (DoS) attacks may also be launched using malware.

Because of the volume of confidential material they keep and the sophistication of their systems, businesses are especially susceptible to malware. Malware may use security flaws to access systems, giving attackers access to private information and harm such systems. Additionally, malware may be used to initiate cyberattacks against a company, interrupting operations and costing money.

Malware is also a threat to businesses since it can swiftly infect networks. Once one computer is compromised, the infection may swiftly spread to more systems, giving hackers access to private information and enabling them to conduct cyberattacks. Additionally, malware may be difficult to find and get rid of, which raises the danger for businesses. Because it may be used to steal sensitive information, interfere with business processes, and conduct cyberattacks, malware poses a severe danger to enterprises. Employing security measures like firewalls, antivirus software, and patch management can help organizations safeguard themselves against malware. Organizations should also teach staff members how to spot possible dangers and take precautions to safeguard their systems [12]. Organizations may lower the risk of a malware attack and safeguard their data and systems by adopting these precautions.

**Ransomware Threat**

A kind of cyberattack known as ransomware encrypts data belonging to an organization and keeps it hostage until a ransom is paid. Hackers may employ a pernicious kind of malware to demand money from their targets [13]. The attacker will often demand a ransom payment for a decryption key that will let the victim recover access to their data.

Because they may result in considerable data loss, business interruption, and reputational harm, ransomware assaults can be fatal for enterprises. The organization may only be aware that they are infected in many circumstances once it is too late. If the ransom is not paid, the attacker could additionally threaten to release the victims' personal information. Malicious emails or links are the most popular means of spreading ransomware when the receiver is duped into clicking on a malicious link or file [12]. Ransomware is often disseminated via drive-by downloads and malicious websites. The ransomware may encrypt the victim's data as soon as the malicious code is downloaded, rendering it unavailable.

Organizations must take precautions to guard against ransomware attacks. They should ensure that all computers have installed the most recent antivirus and security updates and that users know the dangers of opening dubious links and files [13]. Additionally, businesses should frequently back up their data to recover it in the case of a ransomware attack.

**Phishing threat**

Over the last 10 years, phishing has become a more prevalent cyberattack. Attackers aggressively try to obtain a business's sensitive data, such as passwords, banking information, and other private data. Phishing attacks are often carried out by email or instant messaging, as they usually include the attacker delivering a link to what looks to be a trustworthy website or email but is harmful. An organization's sensitive information is the main target of a phishing assault. Attackers may use the info they have access to steal data, log into other accounts, or carry out other nefarious deeds. Additionally, hackers could use the data to carry out other cyberattacks, such as ransomware or distributed denial-of-service (DDoS) operations [14].

Due to a lack of security measures, businesses are especially susceptible to phishing attempts. Businesses often rely on employees to identify possible phishing attacks and take the required precautions to protect themselves [12]. Employees may be unable to see the warning indications of a phishing attempt since attackers are becoming more skilled and adept at creating convincing emails. By training staff about phishing scams, establishing two-factor authentication, and using security software to identify and prevent bad emails, businesses may take precautions to safeguard themselves. Organizations should also consider implementing a system that analyzes staff emails and alerts them to any questionable behavior. Overall, phishing is a significant online danger that impacts businesses of all kinds. Businesses need to take preventative efforts to defend themselves against this kind of assault by training their staff, putting security measures in place, and keeping an eye on their networks for unusual behavior.

**Social Engineering Threat**

Social engineering is a cyberattack where the attacker manipulates the victim's mind to access private data or resources. This one is one of the most prevalent and effective cyber threats facing corporations. Attacks against social engineering often try to take advantage of people's inclination to trust and cooperate. Most of the time, the attacker will utilize trickery, deceit, or informational or technological manipulation to obtain the victim's private information. Phishing emails, malicious websites, telemarketing calls, text messages, and other communication tools may be used by attackers to deceive a victim into divulging personal data or granting access to sensitive resources.

Because they often have insufficient security mechanisms to safeguard their internal resources, organizations are especially susceptible to social engineering assaults. Attackers may use social engineering to acquire private information such as consumer data, financial records, and employee credentials. Attackers may also enter a company's networks and systems through social engineering, which increases the risk of data breaches and other detrimental cyber disasters [14].

Since social engineering assaults depend on human connection, they are extremely difficult to spot. Without the victim's knowledge, attackers can persuade victims to provide access or information. Without the organizations' knowledge, this may expose them to attacks over extended periods. Organizations need to take precautions to guard against social engineering assaults. This entails informing staff members of the risks associated with social engineering, putting in place stringent security measures, and using reliable authentication procedures. In order to recognize and prevent social engineering attacks, organizations also need to adopt cutting-edge security technology like identity and access management systems.

**Denial of Service (DoS) Threats**

A cyberattack, known as a denial of service (DoS), uses excessive malicious traffic to prevent a network, system, or service from operating normally. A DoS attack aims to overload the system with requests so that it cannot react to valid service requests. This may cause the machine to shut down or significantly degrade its performance. DoS attacks may significantly affect enterprises because they can prevent the availability of essential services, leading to lost revenue and unhappy customers. DoS assaults can also divert attention away from other nefarious actions, including data theft or ransomware operations, while the attack is in progress. DoS attacks may also be used to convey a message or to interfere with an organization's activities.

DoS attacks may be executed in several different methods. They may be started from a single computer using a simple script or with a botnet, a decentralized network of machines used to deliver copious volumes of malicious traffic. In order to amplify the effect of the assault, more advanced methods might be utilized, such as DNS amplification attacks and application layer attacks. Employing firewalls, intrusion prevention systems, and other security measures are some actions organizations may take to defend themselves against DoS assaults. However, enterprises must be vigilant in their security procedures due to the sophistication and complexity of DoS assaults. This entails monitoring and analyzing network activity, patching systems often, and putting in place a thorough security plan. DoS attacks pose a significant risk to businesses since they may cause service interruptions, financial losses, and reputational harm [14]. Companies must be aware of the dangers of DoS attacks and take preventative measures to safeguard their networks.

**Data Breach Threat**

When a hacker or unauthorized individual has access to private, sensitive, or otherwise protected data, this is a data breach. This may comprise private company information, financial information, or personal data. Because they may result in identity theft, financial losses, and other catastrophic repercussions, data breaches pose a significant danger to the internet. Data breaches may happen in many different ways, including via malware, phishing emails, or compromised credentials. The hostile actor may use the data to steal money or conduct fraud if they get access to it. They could use it for more sinister reasons or sell it on the dark web.

Data breaches may have a negative effect on both organizations and people. It may cause financial losses, reputational harm, and a loss of consumer confidence for firms. Identity theft, money losses, and other forms of fraud may happen to people. Because malevolent actors may access personal data, it can potentially result in a loss of privacy. Several security measures may be put in place to avoid data breaches. Strong password usage, two-factor authentication, encryption, and vulnerability scan audits are a few of these. Organizations should have a strong incident response strategy to react promptly to any data breaches. In conclusion, data breaches are a severe cyber danger since they may negatively affect both people and enterprises. Organizations should deploy a wide range of security measures and implement an incident response strategy to avoid data breaches.

**SQL Injection Threat**

Web applications that employ SQL databases are the subject of the SQL injection cyberattack. An attacker can insert malicious SQL code into a website and utilize the invalid code to access sensitive data like user passwords, money, and private documents. Because an attacker may access a website or program without having to interact with the system directly, SQL Injection attacks are extremely hazardous. Because it may be exploited to access critical data, impede company operations, or even undermine the whole system, SQL Injection poses a severe security risk to enterprises. By manipulating the SQL code, an attacker may get past application security and access the underlying database. They are then able to see, edit, or remove private information. Using SQL Injection by an attacker also allows for the execution of malicious instructions on the server, such as creating new user accounts or disabling security measures.

Due to old or improperly built web applications, businesses are especially susceptible to SQL Injection attacks. Misusing user input without adequate validation often results in SQL Injection vulnerabilities. This enables the injection of malicious code into the program, which the application and the underlying database will execute. Additionally, organizations may be vulnerable to SQL Injection attacks if they use shoddy authentication methods, such as using default credentials [15]. Organizations must take precautions to guard against SQL Injection attacks. This entails implementing access control procedures, patching any known vulnerabilities, and restricting the number of users who can access the system. Additionally, organizations must ensure that security procedures are frequently reviewed and updated and that application programs are configured correctly. To avoid injecting malicious code into the application, organizations should make sure that the user input method is able.

**Distributed Denial of Service (DDoS) Threat**

A sort of cyber assault called a distributed denial of service (DDoS) uses a lot of malicious traffic or requests to overload a target system. This results in a denial of service for the target since the system cannot reply to valid requests. Any system, including web applications, mail servers, application servers, and even whole networks, is vulnerable to DDoS assaults [16]. DDoS assaults are often launched via a botnet, a collection of infected computers. Malicious actors will use the botnet to send a significant amount of data or queries to the target system. This might result in the target system needing help to handle valid requests. Additionally, the malicious traffic may make the system unstable or slow down, resulting in an even worse denial of service.

DDoS assaults may significantly affect enterprises since they can result in outages, sluggishness, and data loss. DDoS assaults may also divert businesses' attention from more sinister actions like data theft or ransomware operations. Due to the wide range of attack methods that criminal actors might employ, DDoS assaults can be exceedingly challenging to defend against [15]. By deploying suitable network security measures, such as firewalls and DDoS protection services, organizations may reduce the danger of DDoS assaults.

In conclusion, Distributed Denial of Service (DDoS) is a cyber threat that may seriously harm businesses. DDoS assaults have the potential to disrupt operations, slow down systems, leak data, and even serve as a decoy for groups engaged in more sinister actions. By putting in place the right security measures, such as firewalls and DDoS protection services, organizations can lessen the danger of DDoS assaults [16].

**Password Attacks Threat**

Cyber threats like password assaults impact all sorts of companies. In a password attack, a hacker tries to enter a company's system by figuring out or getting the user's password. Since passwords are a systematic method of authentication used to access sensitive information, this kind of attack poses a particularly severe danger. For several reasons, password assaults pose a danger to an organization's security. First, phishing emails and social engineering techniques may often be used to guess or get passwords [17]. To guess passwords, attackers may also utilize automated methods like brute strength or dictionary assaults. In addition, hackers may employ malware or keyloggers to record user keystrokes or steal passwords.

Organizations may suffer major repercussions as a result of password assaults. An attacker might be able to obtain private data or interfere with operations if they manage to get into a system. Additionally, they could implant malware and produce or edit data. Data loss, identity fraud, or financial losses may result from this. Organizations may take precautions to guard against password assaults. This entails employing two-factor authentication, creating complex passwords that are challenging to guess, and changing passwords often.

Additionally, businesses need to educate their users about the value of strong passwords and the dangers of password hacks. Password assaults are a significant cyber danger that may have substantial repercussions for enterprises, to sum up. Organizations should take precautions to defend against these attacks and inform users of the value of strong passwords.

## CONCLUSION

Organizational cyber threats are a serious issue that should not be ignored. A successful cyber-attack can have devastating repercussions, including data loss, reputational harm, financial loss, and other grave ramifications. Organizations should always take the necessary actions to safeguard themselves against these dangers, such as creating a thorough cybersecurity strategy, educating staff on cybersecurity best practices, and putting in place strong security measures. As threats change and new technologies are created, cybersecurity is a process that needs to be continuously updated. Organizations should keep making investments in the tools and technologies required to safeguard their systems, data, and networks against harmful cyber threats.

## REFERENCES

1. Aparajit S, Shah R, Chopdekar R, Patil R. Data Protection: The Cloud Security Perspective. In 2022 IEEE 3rd International Conference for Emerging Technology (INCET). 2022 May 27; 1–5.
2. Ibrokhimov S, Hui KL, Al-Absi AA, Sain M. Multi-factor authentication in cyber physical system: A state of art survey. In 2019 IEEE 21st international conference on advanced communication technology (ICACT). 2019 Feb 17; 279–284.
3. Isaak J, Hanna MJ. User data privacy: Facebook, Cambridge Analytica, and privacy protection. Computer. 2018 Aug 14; 51(8): 56–9.
4. Lin F, Zhou Y, An X, You I, Choo KK. Fair resource allocation in an intrusion-detection system for edge computing: Ensuring the security of Internet of Things devices. IEEE Consum Electron Mag. 2018 Oct 5; 7(6): 45–50.
5. Abdulghani HA, Nijdam NA, Collen A, Konstantas D. A study on security and privacy guidelines, countermeasures, threats: IoT data at rest perspective. Symmetry. 2019 Jun 10; 11(6): 774.

6. El Ghamry M, Halim IT, Bahaa-Eldin AM. Secular: A decentralized blockchain-based data privacy-preserving model training platform. In 2021 IEEE International Mobile, Intelligent, and Ubiquitous Computing Conference (MIUCC). 2021 May 26; 357–363.
7. Alkadi O, Moustafa N, Turnbull B, Choo KK. A deep blockchain framework-enabled collaborative intrusion detection for protecting IoT and cloud networks. IEEE Internet Things J. 2020 May 22; 8(12): 9463–72.
8. Nguyen DC, Pathirana PN, Ding M, Seneviratne A. Blockchain for secure EHRS sharing of mobile cloud based e-health systems. IEEE Access. 2019 May 17; 7: 66792–806.
9. Daoud L, Huen H. Performance Study of Software-based Encrypting Data at Rest. Proceedings of 37th International Confer. 2022 Mar 18; 82: 122–30.
10. Sun P. Security and privacy protection in cloud computing: Discussions and challenges. J Netw Comput Appl. 2020 Jun 15; 160: 102642.
11. Mavroeidis V, Hohimer R, Casey T, Jesang A. Threat actor type inference and characterization within cyber threat intelligence. In 2021 IEEE 13th International Conference on Cyber Conflict (CyCon). 2021 May 25; 327–352.
12. Gao Y, Xiaoyong LI, Hao PE, Fang B, Yu P. Hincti: A cyber threat intelligence modeling and identification system based on heterogeneous information network. IEEE Trans Knowl Data Eng. 2020 Apr 20; 34(2): 708–722.
13. Williams R, Samtani S, Patton M, Chen H. Incremental hacker forum exploit collection and classification for proactive cyber threat intelligence: An exploratory study. In 2018 IEEE International Conference on Intelligence and Security Informatics (ISI). 2018 Nov 9; 94–99.
14. Abu MS, Selamat SR, Ariffin A, Yusof R. Cyber threat intelligence–issue and challenges. Indones J Electr Eng Comput Sci. 2018 Apr; 10(1): 371–9.
15. Jothi KR, Pandey N, Beriwal P, Amarajan A. An efficient SQL injection detection system using deep learning. In 2021 IEEE International Conference on Computational Intelligence and Knowledge Economy (ICCIKE). 2021 Mar 17; 442–445.
16. Dong S, Abbas K, Jain R. A survey on distributed denial of service (DDoS) attacks in SDN and cloud computing environments. IEEE Access. 2019 Jun 12; 7: 80813–28.
17. Badsha S, Vakilinia I, Sengupta S. Privacy preserving cyber threat information sharing and learning for cyber defense. In 2019 IEEE 9th Annual Computing and Communication Workshop and Conference (CCWC). 2019 Jan 7; 0708–0714.